\documentclass[twocolumn]{aastex62}

\usepackage{subfigure}
\usepackage{booktabs}
\usepackage{multirow}
\usepackage{color,soul}
\newcommand{\hi}{\ion{H}{1}}
\newcommand{\cvi}{\ion{C}{6}}
\newcommand{\nvi}{\ion{N}{6}}
\newcommand{\nvii}{\ion{N}{7}}
\newcommand{\ovii}{\ion{O}{7}}
\newcommand{\oviii}{\ion{O}{8}}

\newcommand{\neix}{\ion{Ne}{9}}
\newcommand{\nex}{\ion{Ne}{10}}
\newcommand{\mgxi}{\ion{Mg}{11}}

\def\hi{{{\rm H}\,{\sc i}}}
\def\cvi{{{\rm C}\,{\sc vi}~}}
\def\nvi{{{\rm N}\,{\sc vi}~}}
\def\nvii{{{\rm N}\,{\sc vii}~}}
\def\ovii{{{\rm O}\,{\sc vii}~}}
\def\oviii{{{\rm O}\,{\sc viii}~}}

\def\neix{{{\rm Ne}\,{\sc ix}~}}
\def\nex{{{\rm Ne}\,{\sc x}~}}
\def\mgxi{{{\rm Mg}\,{\sc xi}~}}

\def\suzaku{{\it Suzaku}}
\def\chandra{{\it Chandra}}
\def\xmm{{\it XMM-Newton}~}
\def\rosat{{\it ROSAT}~}

\shorttitle{Multi-thermal hot CGM}
\shortauthors{Das et al.}

\begin{document}

\title{Multiple temperature components of the hot circumgalactic medium of the Milky Way}

\correspondingauthor{Sanskriti Das}
\email{das.244@buckeyemail.osu.edu}

\author[0000-0002-9069-7061]{Sanskriti Das}
\affiliation{Department of Astronomy, The Ohio State University, 140 West 18th Avenue, Columbus, OH 43210, USA}

\author{Smita Mathur}
\affiliation{Department of Astronomy, The Ohio State University, 140 West 18th Avenue, Columbus, OH 43210, USA}
\affil{Center for Cosmology and Astroparticle Physics, 191 West Woodruff Avenue, Columbus, OH 43210, USA}

\author{Anjali Gupta}
\affiliation{Department of Astronomy, The Ohio State University, 140 West 18th Avenue, Columbus, OH 43210, USA}
\affil{Columbus State Community College, 550 E Spring St., Columbus, OH 43210, USA}

\author{Fabrizio Nicastro}
\affiliation{Observatorio Astronomico di Roma - INAF, Via di Frascati 33, 1-00040 Monte Porzio Catone, RM, Italy}
\affiliation{Harvard-Smithsonian Center for Astrophysics, 60 Garden St., MS-04, Cambridge, MA 02138, USA}
\author{Yair Krongold}
\affiliation{Instituto de Astronomia, Universidad Nacional Autonoma de Mexico, 04510 Mexico City, Mexico}



\begin{abstract}
\noindent We present a deep \xmm observation of the Galactic halo emission in the direction of the blazar 1ES\,1553+113. In order to extract the Galactic halo component from the diffuse soft X-ray emission spectrum, accurately modeling the foreground components is crucial. Here we present  complex modeling of the foregrounds with unprecedented details. 
 A careful analysis of the spectrum yields two temperature components of the halo gas (T$^{em}_1$= \textcolor{black}{10$^{6.25-6.42}$K}, T$^{em}_2$= \textcolor{black}{10$^{6.68-6.92}$K}). We find that these temperatures obtained from the emission spectrum are not consistent with those from the absorption spectrum (T$^{ab}_1$= \textcolor{black}{10$^{6.07-6.13}$K}, T$^{ab}_2$= \textcolor{black}{10$^{6.96-7.15}$K}), unlike the previous studies that found only one temperature component of the Milky Way circumgalactic medium. This provides us with interesting insights into the nature of emitting and absorbing systems. We discuss several possibilities objectively, and conclude that most likely we are observing multiple (3 to 4) discrete temperatures between 10$^{5.5}$K and $\geqslant$10$^7$K \textcolor{black}{in the Milky Way circumgalactic medium}.     
\end{abstract}



\section{Introduction} \label{sec:intro}
\begin{figure*}
\centering
\includegraphics[scale= 0.5]{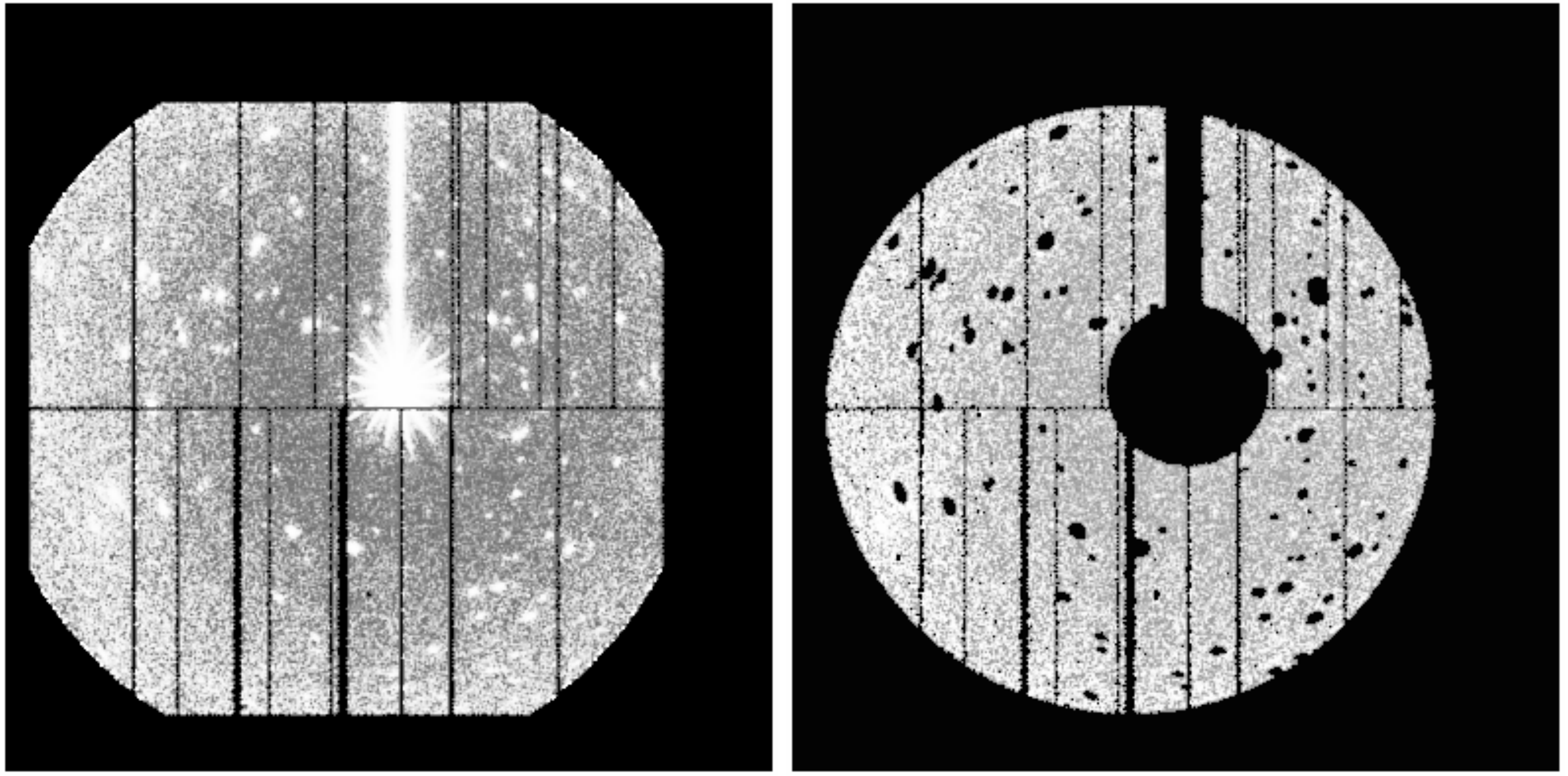}
\caption{The \textcolor{black}{exposure-corrected} EPIC-pn smoothed images in the detector coordinate before (left) and after (right) point source subtraction, masking the target blazar 1ES\,1553+113 and its readout streak, and removing the corners. Note that the images are not vignetting corrected; the instrumental background and the soft proton contamination have not been subtracted off either. The outer and inner radii of the annuli in the right panel are 12.6$'$ and 3.4$'$, respectively.}
\label{fig:Image}
\end{figure*}
\noindent The circumgalactic medium (CGM) is the halo of multi-phase gas and dust surrounding the stellar disk and the interstellar medium of a galaxy, within the virial radius \citep{Putman2012,Tumlinson2017}\footnote{\textcolor{black}{Historically, the circumgalactic gas of the Milky Way has been referred as the Galactic ``halo" or ``corona". CGM is a more prevalent term for external galaxies. However, as they are essentially the same thing, we will use these terms interchangeably.}}. It is a very important component of a galaxy harboring a large fraction of its missing baryons and missing metals \citep{Gupta2012,Peeples2014}. Numerical simulations show that the properties of the CGM are governed by galaxy mass, and are affected by accretion from the intergalactic medium (IGM) and feedback from the galactic disk \citep{Ford2014,Oppenheimer2016,Oppenheimer2018}. Precipitation from the CGM in turn may help sustain next generation of star-formation in a galaxy \citep{Voit2015}. Thus the CGM plays an important role in the evolution of a galaxy. 

The CGM is multiphase in its ionization states, spanning over two orders of magnitude of temperature: T $\approx$ 10$^{4-6}$ K \citep{Ford2014,Oppenheimer2016,Nicastro2016a,Suresh2017}. The hot (T $\approx$ 10$^6$ K) gaseous Galactic corona at the virial temperature has been a long-standing prediction \citep{Spitzer1956} and is believed to be the most massive component of the CGM. This phase can be probed by highly ionized metals (e.g. oxygen). The dominant transitions of oxygen, \ovii and \oviii lie in the soft X-ray band. The distribution of the phase structure (density and temperature), metallicity, kinematics, the spatial extent, and the mass of this hot gas provide important constraints to the models of galaxy formation, the accretion and feedback mechanisms, and the co-evolution of the CGM with the galaxies. 

The search for hot gas beyond the optical radii of galaxies started with \rosat and continued with \chandra, \xmm and \suzaku. In this paper, we focus on the hot CGM of the Milky Way. Because of our special vantage point, the highly ionized CGM of the Milky Way has been studied in emission \citep{Snowden2000,Kuntz2000,Henley2010,Henley2013,Henley2015a,Henley2015b,Nakashima2018} and absorption \citep{Gupta2012,Gupta2014,Nicastro2016a,Nicastro2016b,Gupta2017,Nevalainen2017,Gatuzz2018} in much better details compared to other galaxies. The combined studies of emission and absorption have shown that the CGM is diffuse, warm-hot (T $\approx$ 10$^{6.3}$ K), extended, and massive \citep{Gupta2012,Nicastro2016b}, as well as anisotrpic \citep{Henley2010,Gupta2014,Gupta2017}. Similar studies along more sightlines will provide a larger solid angle coverage and a more complete picture of the CGM characteristics, informing the theories of galaxy evolution. 

By analyzing a very deep (t$_{exp}$ = 1.85 Ms) absorption spectrum in the sightline of the blazar 1ES\,1553+113 observed with \xmm RGS (Reflection Grating Spectrometer), \citet{Das2019a} discovered a 10$^7$ K CGM component coexisting with the warm-hot 10$^6$ K CGM. Here we have expanded on their work by studying the emission spectrum around the same sightline for a better understanding of the multi-component highly ionized CGM. Inspired by the two-temperature hot CGM obtained from their absorption studies along this sightline, we have modelled the emission spectrum with a two-temperature Galactic halo. The aim was to obtain the density and the spatial extent of the observed gas by combining the emission and absorption measurements, as has been done previously for one-temperature halo model \citep{Gupta2012,Gupta2017}. Interestingly, however, we have found that the temperatures of the emitting and the absorbing gas are not the same, unlike the previous studies (discussed in \S\ref{compare:ab}). This indicates that we are not observing the same gas in emission and absorption. This also shows us that the highly ionized halo gas consists of at least three (or four) components, rather than the two components discovered earlier. A single component at the virial temperature is clearly ruled out; this is discussed in section \ref{sec:physics}.    

Our paper is structured as follows: we discuss the data reduction and analysis in section \ref{sec:reductionandanalysis} and discuss our results in section \ref{sec:results}. We interpret the results, and compare them with existing emission and absorption studies in section \ref{sec:discussion}. Finally, we summarize our results and outline some of the future aims in section \ref{sec:conclusion}.
\section{Data reduction and analysis}\label{sec:reductionandanalysis}
\noindent Our goal is to extract and analyze the diffuse X-ray emission spectrum surrounding the 1ES\,1553+113 sightline, observed with \xmm EPIC-pn and MOS in PrimeFullWindow mode.  The details of the observations are presented in \citet{Nicastro2018} and \citet{Das2019a}. 
\subsection{Data reduction}\label{sec:reduction}
\noindent The total observation has 18 exposures, but in this analysis we do not use 5 datasets\footnote{ObsID: 0094380801, 0656990101, 0727780101, 0727780201, 0727780301} with low signal-to-noise ratio (S/N). We have reduced the data of 13 observations\footnote{ObsID:0761100101, 0761100201,0761100301, 0761100401, 0761100701, 0761101001, 0790380501, 0790380601, 0790380801, 0790380901, 0790381001, 0790381401, 0790381501}
using XMM-Extended Source Analysis Software (ESAS)\footnote{\url{ftp://xmm.esac.esa.int/pub/xmm-esas/xmm-esas.pdf}}. The total exposure time is 1.665 Ms. We follow the standard procedure of filtering, point source identification and removal, spectra and detector background extraction using default conditions except the ones explicitly mentioned here. {Each observation is processed separately}. In the point source detection routine \texttt{cheese}, we tune the following parameters: 1) The PSF threshold parameter \textit{scale} is changed to 0.20 from 0.25; this allows us to remove a larger fraction of the point source flux.  2) Minimum separation for point sources \textit{dist} is changed to 15$''$ from 40$''$. This allows us to detect close-by sources. The Epic-pn PSF is 12.5$''$ (FWHM), thus we ensure that all the resolved sources are counted.  3) The point-source flux threshold \textit{rate} is changed to 0.01 from 1.0 (in the unit of 10$^{-14}$ ergs cm$^{-2}$ s$^{-1}$). This ensures that we identify and remove fainter sources (figure \ref{fig:Image}). Additionally, we remove a circular region of $\approx$3.4$'$ radius around 1ES\,1553+113 to mask out the blazar (figure \ref{fig:Image})\footnote{{The encircled energy fraction of EPIC-pn is $\approx$1 at an angular radius of 2.5$'$. Therefore, the masked out region of 3.4$'$ radius certainly removes any stray light from the blazar. This confirms that the emissions we detect are not contaminated by the blazar.}}. {Each observation is carefully checked after source removal to make sure that any visibly identifiable source is not present. Also, we make sure that the set of identified sources in all observations is similar.} The readout streaks from the very bright sources, if any, are removed (figure \ref{fig:Image}). The spectra are extracted from a circular region of $\approx$0.21 deg to avoid the excess soft proton contamination in the corners of the CCDs, as recommended in the ESAS manual. Two observations (0761100701, 0790381001) did not have any data from EPIC-pn and one observation (0790380901) did not have out-of-time (OoT) information; we do not use these observations further. All the other pn spectra are OoT subtracted.  The effective exposure time of the 10 observations after filtering is $\approx$ 807 ks, $\sim$60\% of the unfiltered exposure time (1.339 Ms). Each spectrum is binned using ftool \textit{grppha} such that minimum count in each bin is 50, which gives a moderate S/N.   
\subsection{Data analysis}\label{sec:analysis}
\begin{figure*}. 
    \centering
    \includegraphics[trim= 0 0 40 0,clip,scale=0.65]{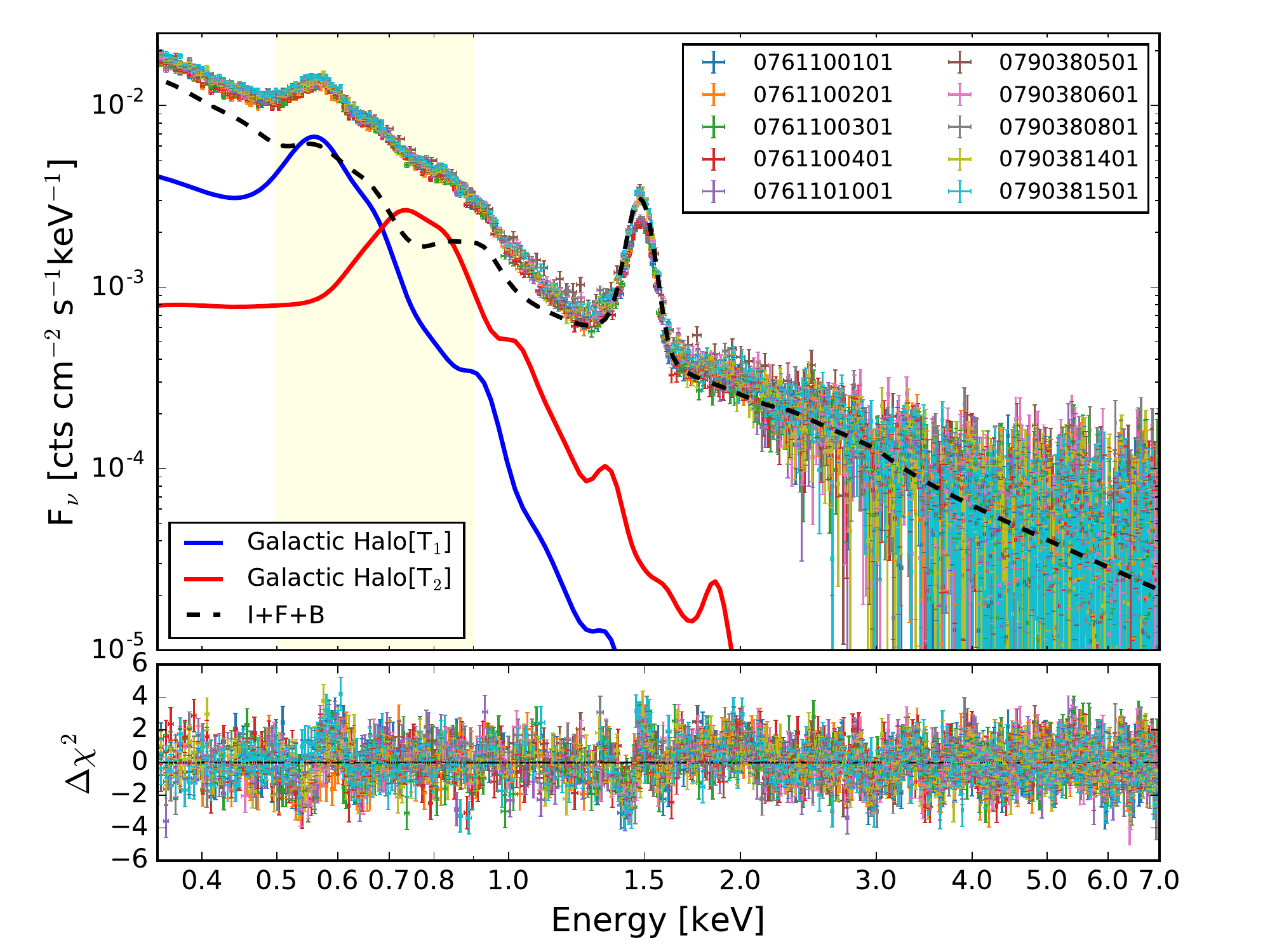}
    \includegraphics[trim= 0 0 40 0,clip,scale=0.65]{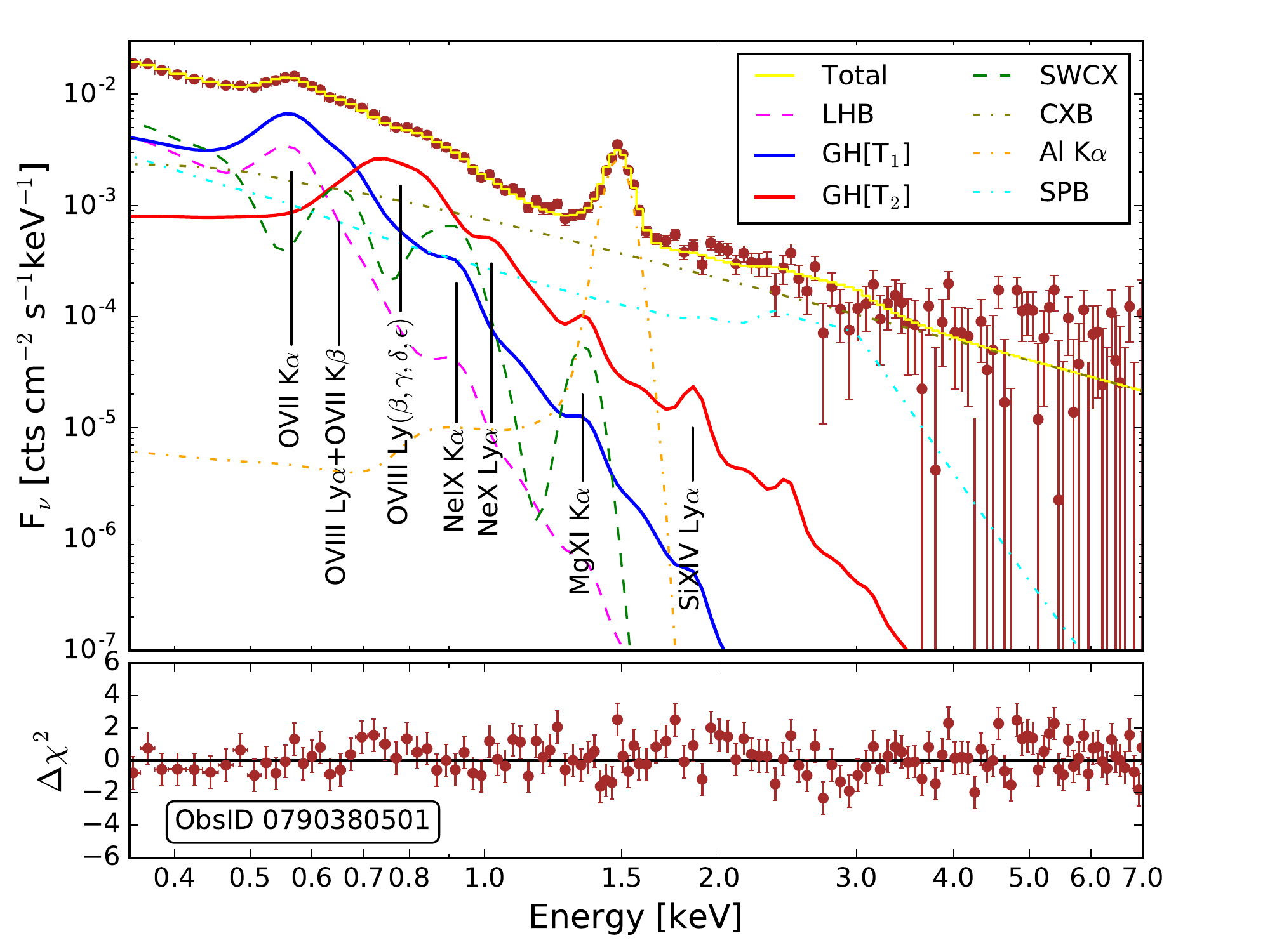}
    \caption{Top: The whole set of spectra with the best-fitted model. All instrumental, foreground and background components have been added to show their cumulative contribution in the spectra (in black dashed line). Galactic halo components are shown in blue (warm-hot) and red (hot). The vertical yellow strip marks the sweet spot of detecting the halo emission at different temperatures, beyond which the foreground (at lower energy) or the background (at higher energy) dominates. Bottom: One spectrum is decomposed into the instrumental line, foreground (model F), background and the halo (see appendix \ref{sec:model} for the details of the legend). The characteristic emission lines of oxygen, neon, magnesium and silicon are labeled, so that the relative contribution of the spectral components at each line can be studied independently. For example, the warm-hot (shown in blue) system emits most of the \ovii K$\alpha$, while most of the neon and \oviii emission come from the SWCX and the hot (shown in red) phase.}
    \label{fig:pn}
\end{figure*}
\noindent We have done all the spectral analysis on pn spectra, as it has the larger effective area, therefore larger photon count. We have also analyzed the MOS2 spectra to check for consistency (see appendix \ref{sec:mos}), because its spectral resolution is better than pn. As MOS1 has only 3 usable CCDs, we do not use it in our analysis. 

Extracting the Galactic halo component from the diffuse X-ray spectrum is notoriously difficult \citep{Henley2015b} as the spectrum is composed of four different components: 1) instrumental lines and soft proton contamination, 2) foreground, 3) background, and 4) the Galactic halo. The foreground itself is made up of two different components, the solar wind charge-exchange (SWCX) and the local hot bubble (LHB). Due to the large uncertainty in the nature of the foreground component(s), we consider 6 foreground models to see how they affect our measurement of the Galactic halo. This allows us to estimate the systematic uncertainty due to foreground modeling. Our model of the Galactic halo also differs from most of the earlier models. As \cite{Das2019a} found two distinct temperatures in the highly ionized absorption systems along the same sightline, we fit the Galactic halo emission spectrum with two temperature components. We allow both the components to vary in the same temperature range of 10$^{5.5-7.5}$ K, and do not force them to be different. Because of the complexity of the spectral modeling, we have been extremely careful and conservative in our analysis. The details of the spectral model(s) are discussed in appendix \ref{sec:model}. 

In our analyses, we assume all plasma components to be in collisional ionization equilibrium (CIE). The cosmological hydrodynamic simulations \citep{Dave2010} and the high resolution simulations focused on individual galaxies \citep{Stinson2012} show that the CGM is in CIE. Previous X-ray observations \citep{Henley2010,Gupta2012,Henley2013,Gatuzz2018} are also consistent with the plasma being in CIE, validating our assumption.

While calculating the error bars of any parameter using \texttt{err}, we do not freeze other parameters which were not frozen during the fit; this ensures that the errors are not underestimated. If $\chi^2$ is non-monotonic with respect to any parameter \citep[also seen by][]{Henley2015a}, we derive the confidence interval and the error bars of that parameter using \texttt{steppar}. The best-fitted values from the 6 foreground models (appendix \ref{sec:model}) add a systematic uncertainty to each fitted parameter on top of the statistical uncertainty of that parameter (table \ref{table:param}) in each model. The quoted best-fitted values of the halo parameters are the average of their respective best-fitted values in the 6 models. We make sure that the mean is not biased towards any particular model by comparing it with the median. The corresponding statistical uncertainties have been propagated in averaged quadrature. For each measured parameter, we quote the best-fitted value $\pm$ statistical uncertainty $\pm$ systematic uncertainty. Unless explicitly mentioned otherwise, we quote uncertainties as 1$\sigma$ error bars, and ranges as 68\% confidence intervals.   

\section{Results}\label{sec:results}
\begin{table*}[]
\centering
\caption{The best-fitted values of the halo parameters for 6 foreground models (see appendix \ref{sec:model} for details). Errors of the normalization are 1$\sigma$, while the errors in temperature are 3$\sigma$; these are statistical errors only. }
\label{table:param}
\begin{tabular}{cccccc}
       & \multicolumn{2}{c}{Warm-hot}                                       & \multicolumn{2}{c}{Hot}             \\ \cmidrule(l){2-5} 
Models & k$_B$T (keV) & Norm ($\times$10$^{-4}$cm$^{-5}$)   & k$_B$T (keV)   & Norm ($\times$10$^{-4}$cm$^{-5}$) & $\chi^2/dof$  \\ \midrule
A      & 0.226$\pm$0.007           & 7.70$^{+0.06}_{-0.08}$  & 0.708$^{+0.032}_{-0.026}$ & 1.09$\pm0.05$   & 13102.47/12259      \\ 
B      & 0.187$\pm$0.001           & 12.81$^{+0.02}_{-0.04}$ & 0.644$\pm$0.010           & 1.77$\pm0.01$    & 12898.84/12257     \\ 
C      & 0.155$^{+0.018}_{-0.024}$ & 10.76$^{+1.39}_{-0.38}$ & 0.467$^{+0.033}_{-0.067}$ & 2.35$\pm0.01$     & 12684.26/12252    \\
D      & 0.208$\pm$0.001           & 9.24$^{+0.63}_{-0.18}$  & 0.693$^{+0.010}_{-0.017}$ & 1.47$\pm0.01$      & 12903.57/12256   \\
E      & 0.148$^{+0.020}_{-0.013}$ & 13.94$^{+1.91}_{-0.52}$ & 0.491$^{+0.020}_{-0.050}$ & 2.39$\pm0.01$       & 12735.66/12250  \\
F      & 0.172$^{+0.039}_{-0.045}$ & 8.21$^{+2.69}_{-1.48}$  & 0.432$^{+0.070}_{-0.071}$ & 2.58$^{+0.15}_{-0.34}$ & 12582.10/12252\\ \midrule
\end{tabular}%
\end{table*}

\noindent We show the spectra and the best-fitted model  in figure \ref{fig:pn} (model F; see Appendix \ref{sec:model} for the model description). In the top panel, we combine the instrumental, foreground and background components to show their contribution in the spectra relative to the halo components. The highlighted region is the energy range suitable to search for the halo signals. Given the spectral resolution, the halo cannot be resolved into more than two temperature components. In the bottom panel, we decompose the model to show how the characteristic line transitions are attributed to different spectral components. 

Because of the different $\frac{\hbox{\oviii}}{\hbox{\ovii}}$ predicted by the LHB and the SWCX models (see figure 12 of \cite{Henley2015b}), the temperature of the warm-hot component with different foreground models are not same (table \ref{table:param}, 2nd column). This adds a large systematic uncertainty on top of the statistical uncertainty. On average, we obtain log (T$^{em}_1$/K) = 6.32$^{+0.04}_{-0.05}\pm0.30$ (99.73\% confidence interval). The temperature of the hot component is affected by two factors: the intensity of \oviii lines at $\approx$0.7-0.8 keV, and the presence/inclusion of Ne and Mg lines in the SWCX models. The \oviii Ly$\beta$, Ne and Mg lines are much weaker than other oxygen lines (e.g. \ovii K$\alpha$, \oviii Ly$\alpha$, \ovii K$\beta$) in models A, B, and D. In models C, E, and F, \oviii Ly$\beta$, Ne and Mg lines are not necessarily weak. The noticeable presence/absence of these lines in the SWCX models decrease/increase the temperature of the hotter Galactic halo, adding a large systematic uncertainty (table \ref{table:param}, 4th column). On average, we obtain log (T$^{em}_2$/K) = 6.82$^{+0.02}_{-0.03}$ $^{+0.28}_{-0.40}$ (99.73\% confidence interval). 

Similarly, the derived emission measure (EM = $\int n_p n_e dl$) of both of the Galactic Halo components\footnote{EM is derived from the normalization using d$\Omega/4\pi$ = 3.049 $\times$ 10$^{-6}$. The effective FOV is point-source-subtracted.} depends on the foreground model (table \ref{table:param}, 3rd and 5th columns). This is because the temperature and the amplitude of the plasma emission are strongly correlated for a given line intensity, adding a large systematic error to the EM. The average emission measure of the warm-hot component is EM$_1$ = 1.06$^{+0.06}_{-0.02}$ $^{+0.62}_{-0.35}$ $\times$ 10$^{-2}$ cm$^{-6}$ pc, and for the hot component it is EM$_2$ = 2.19$^{+0.03}_{-0.06}$ $^{+0.71}_{-1.08}$ $\times$ 10$^{-3}$ cm$^{-6}$ pc. 

\begin{figure}
    \centering
    \includegraphics[trim=25 0 50 20, clip, scale=0.5]{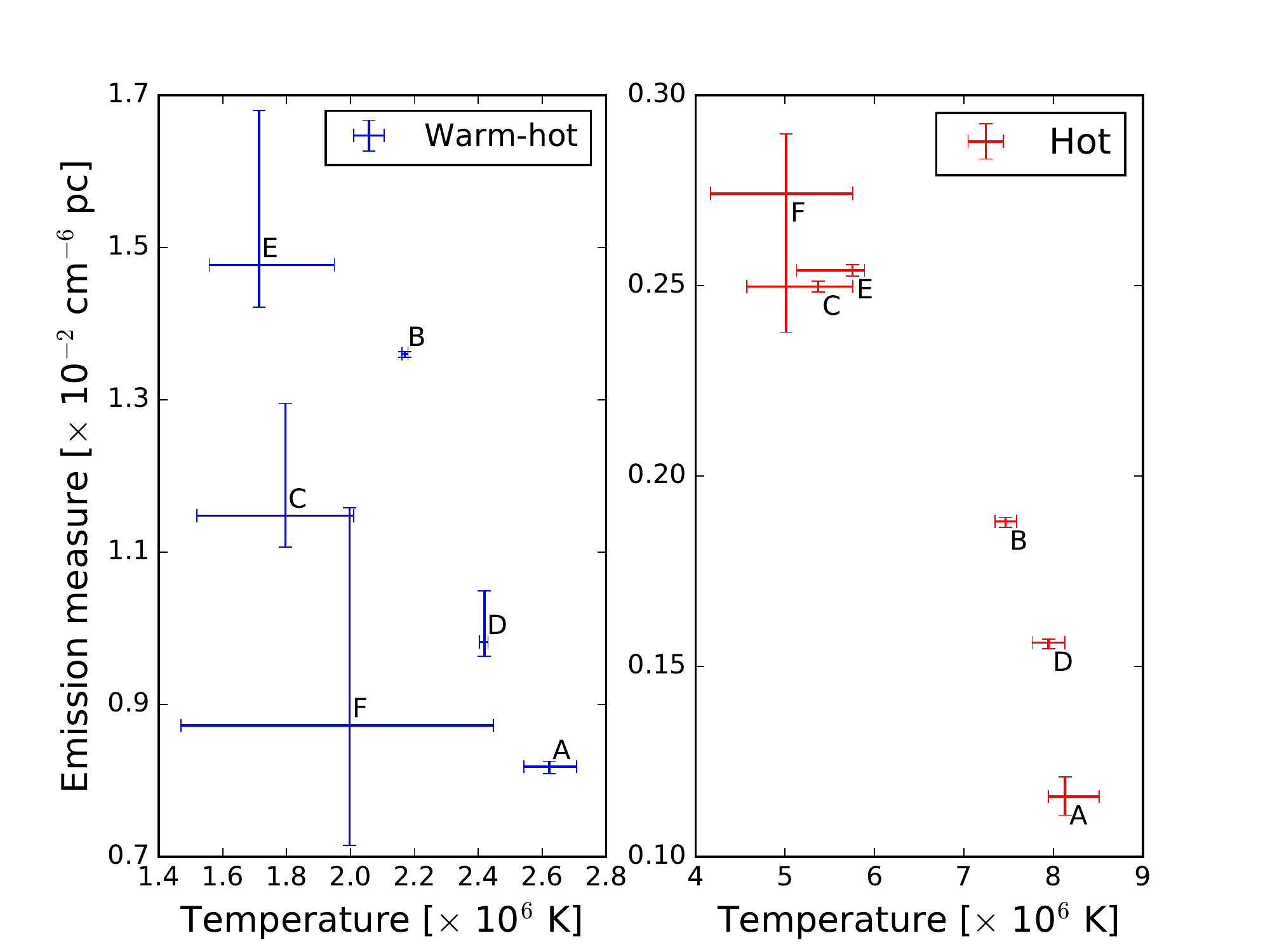}
    \caption{The temperature and the emission measure of the halo components for different foreground models. Each point has been labeled with the corresponding spectral model described in appendix \ref{sec:model}}
    \label{fig:corr}
\end{figure}

The larger temperature of the hot component T$^{em}_2$ produces lower emission measure EM$_2$, because \oviii Ly$\beta$ and higher order lines are stronger at higher temperature (figure \ref{fig:corr}, right panel; table \ref{table:param}, 4th and 5th columns). However, due to the complex interplay between the foreground and the warm-hot component, there is no such trend between the temperature T$^{em}_1$ and the emission measure EM$_1$ of the warm-hot component (figure \ref{fig:corr}, left panel). Interestingly, if the foreground models are split into two groups, 1) models C, E, F (continuum+line foregrounds) and 2) models A, B, D (continuum only foregrounds), a trend similar to the hot component can be seen separately in each group (table \ref{table:param}, 2nd and 3rd columns). In general, continuum only foregrounds predict larger temperatures than continuum$+$line foregrounds at a similar emission measure. However, there is no systematic trend with regards to the nature of foreground component (LHB and/or SWCX), showing how uncertain and complex the foreground modeling is.   

We should note that we cannot choose one model over the other based on $\chi_\nu^2$ value. There is no physical reason to rule out any of these models either. LHB only (model A) and SWCX only (model B, C) are the two extremes of the foreground models, while their combination is a physically better representation of the foreground. Some of the models contain individual lines (e.g. model C, E, F) and some do not. As the fitting routine treats the lines and the continuum differently, considering one model better than the other based on statistic can be spurious. Thus we do not prefer a model, but take into account the foreground variations in the systematic uncertainty of the derived parameters. Figure \ref{fig:corr} shows how the measured temperature and the emission measure of the Galactic halo differ for the different foreground models: about 0.3 dex for T$_1^{em}$ and T$_2^{em}$, 0.4 dex for EM$_1$, and about half a dex for EM$_2$, as noted above. This quantifies the systematic uncertainty.  

\section{Discussion}\label{sec:discussion}
\noindent We have detected two-temperature hot Galactic halo components from the X-ray emission spectra surrounding the 1ES\,1553+113 sightline. Below, we compare our result with similar studies in literature and specifically with the absorption analysis toward the same sightline.
\subsection{Comparison with earlier measurements}\label{compare:em}
\noindent The temperature of the warm-hot component is in excellent agreement with earlier emission and absorption measurements along many other directions \citep{Henley2015a,Gupta2017}, and is consistent with the most likely temperature \citep{Henley2010,Gupta2012,Henley2013,Henley2015b,Nakashima2018} of the warm-hot halo. The emission measure of the warm-hot component is a $\sim$ factor of 2 higher than the average \citep{Henley2010}, but is similar to the emission measure along the Mrk\,509 sightline found by \citet{Gupta2017}, and well within the large range of EM spanning $\sim$ factor of 50 \citep[see][their figure 7, for a comprehensive picture]{Henley2015b}. 

\cite{Kuntz2000} and \cite{Snowden2000} inferred a bimodal temperature distribution of the Galactic halo, based on the spectral characteristics and the angular variation of the soft X-ray emission in \rosat All-Sky Survey data.  Their two temperature model had T$_s$ = 10$^{5.96-6.14}$K and T$_h$ = 10$^{6.42-6.51}$K for the soft and hard components respectively. These are smaller than T$^{em}_1$ and T$^{em}_2$ that we find along the 1ES\,1553+113 sightline, and the differences can be due to several factors including the \rosat energy range,  analysis method and/or the foreground modeling. Their temperatures were determined based on the hardness ratios in the $\frac{3}{4}$ keV and $\frac{1}{4}$ keV bands. On the other hand, we determine the temperature by spectral analysis. Our foreground models are more complete, especially because of the inclusion of the SWCX. Unlike \textit{ROSAT}, our \xmm data are not sensitive below $0.3$ keV, making it difficult for us to detect the T$_s$ component. The T$_h$ component observed by \cite{Kuntz2000} might be an average of the T$^{em}_1$ and T$^{em}_2$ components we find, which could not be resolved due to the poor spectral resolution of \rosat in the $\frac{3}{4}$ keV band. Therefore, the halo gas may comprise of 3 (or more) components at T$_s$, T$^{em}_1$, T$^{em}_2$, as we discuss further in \S\ref{sec:physics}. Alternatively, the different inferred temperatures may simply be the difference between a particular sightline (as discussed here), and the average (as inferred by the all sky survey).     

Ours is the first detection of the two-temperature (hot and warm-hot) halo gas along a sightline; these were not detected previously in emission or absorption along other sightlines. \cite{Nakashima2018} detected a similarly hot halo component in emission along some anti-center directions, but their halo model had only one temperature component. \cite{Henley2013} found an excess emission around 1 keV toward the sightline of NGC 1365 (their sightline 83, figure 2) and on a small number of other sightlines. This led them to fit a 2-T model. One of their temperatures coincides with our warm-hot component T$^{em}_1$, and the other one is at 10$^7$ K. However, they did not identify the source of the 10$^7$ K component as the CGM. Moreover, their foreground models did not have any SWCX component, neither had the foreground models of \cite{Nakashima2018}. Thus it is possible that the excess emission they detect is (at least, partially) a manifestation of strong SWCX emission at energies higher than 0.6 keV, including the \oviii Ly$\beta-\epsilon$ lines. We have performed extensive modeling of the SWCX emission, making sure that the foreground is not underestimated, thus unambiguously detecting both the warm-hot and hot components.  
\subsection{Comparison with the absorption study}\label{compare:ab}
\noindent In the absorption spectrum of 1ES\,1553+113, a 10$^7$ K component was discovered, coexisting with a warm-hot component at $\approx 10^6$ K \citep{Das2019a}. Here, we compare the temperature estimates of the emission and the absorption measurements along the same sightline.
\begin{figure*}
    \centering
    \includegraphics[trim=0 0 0 150, clip, scale=0.75]{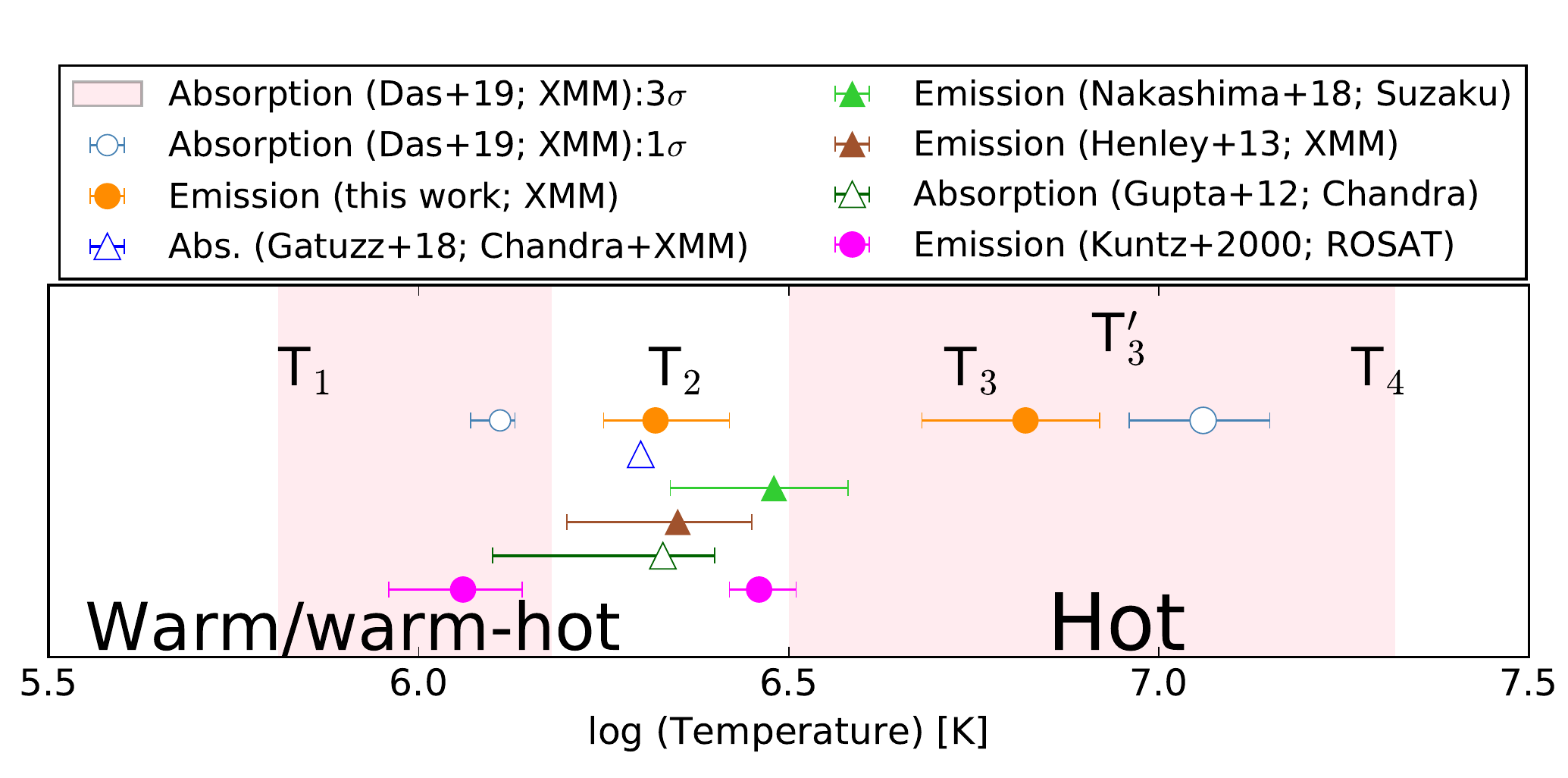}
    \caption{Temperature estimates from emission and absorption studies. The triangles and circles represent 1-T and 2-T CGM model(s), respectively. The filled and unfilled symbols correspond to measurements from emission and absorption studies, respectively. The points from \citet{Kuntz2000,Gupta2012,Henley2013,Nakashima2018,Gatuzz2018} are medians and quartile ranges averaged over many sightlines (without systematic uncertainties), while our work is along one sightline (including systematic uncertainty). Taking into account the different energy coverage of detectors, differences in methodologies and model descriptions, and different spectral aspects of emitting and absorbing components, we infer that the X-ray traced CGM has three (T$_1$, T$_2$, T$'_3$) or four (T$_1$, T$_2$, T$_3$, T$_4$) phases/components.}
    \label{fig:T}
\end{figure*}
\subsubsection{The warm-hot component}\label{sec:warmhot}
\noindent The best-fit temperature of the warm-hot component from the absorption study was log (T$^{ab}_1$/K) = 6.11$^{+0.19}_{-0.49}$ (99.73\% confidence interval). This, however, is smaller than what we find here in emission: log (T$^{em}_1$/K) = 6.32$^{+0.04}_{-0.05}\pm0.30$ (99.73\% confidence interval). This disparity can be due to following reasons: \\
1) \textit{Emissivity bias:}  Absorption strength depends on the number density of ions only. The emission strength, on the other hand, depends both on the number density and emissivity. Therefore, emission is biased towards the gas of higher emissivity. The emissivity of the main tracer \ovii peaks at $\approx 10^{6.3}$ K and decays sharply with temperature in both sides \citep{Yao2009a}. Therefore, the emission spectrum is dominated by the $10^{6.3}$ K component. Even if the $10^{6.1}$ K gas is present, its emissivity would be too low to detect it in emission, given the data quality.  \\
2) \textit{Simplified absorption model:} The absorption spectrum of the 1ES\,1553+113 sightline required a two temperature model discussed in \citet{Das2019a}. It is possible that an additional $10^{6.3}$K thermal component was present, but was not required by the data. Fitting a 3-T halo model in absorption is needed to verify if this is the case. \\
3) \textit{Effect of 2-T modelling:} Earlier studies along many sightlines have found excellent agreement between the temperatures obtained in emission and absorption studies \citep{Henley2010,Gupta2012,Gupta2017}. However, all of those models had only one component for the Galactic halo. As mentioned in \cite{Das2019a}, \oviii was significantly present in the hotter component. Therefore, the assumption of \oviii solely coming from the warm-hot component would overestimate $\frac{\hbox{\oviii}}{\hbox{\ovii}}$, and so the temperature of the warm-hot component. The apparent consistency found earlier might be an artifact of using only oxygen as the tracer in both emission and absorption, which cannot probe beyond one temperature component. Alternatively, the sightline toward 1ES\,1553+113 may be a special case, and the hot component might not be a prevalent component in the CGM of Milky Way.  

While we cannot come to any definite conclusion, it is clear that there are multiple temperature components around 10$^6$ K, which can be probed through a combined study of emission and absorption, using multiple elements as tracers. Most likely we are not seeing the same phase in emission and absorption.  
\subsubsection{The hot component}\label{sec:hot}
\noindent The best-fit temperature of the hot component from the absorbing studies was: log (T$^{ab}_2$/K) = 7.06$^{+0.80}_{-0.72}$ \citep[99.73\% confidence interval,][]{Das2019a}.  This is consistent with the temperature measured from emitting studies (T$^{em}_2$, \S\ref{sec:results}), given the large uncertainty in T$^{ab}_2$. By combining the emission measure EM$_2$(\S\ref{sec:results}) and the equivalent Hydrogen column density N$_{H,2}$= 2.9$^{+2.4}_{-2.0} \times$ 10$^{20}$ cm$^{-2}$\citep{Das2019a}, we find that the average density is n$_{avg}$= EM$_2$/N$_{H,2}$ $\approx$ 1.94$^{+1.48}_{-1.87} \times$ 10$^{-5}$ cm$^{-3}$ and path length is L = $\frac{(N_{H,2})^2}{EM_2}$ = 4839$^{+11572}_{-4645}$ kpc. The errors in density and the path length are large due to the huge uncertainty in N$_{H,2}$. At a conservative limit, we can claim that n$_{avg} \leqslant \frac{max(EM_2)}{min(N_{H,2})}$ = 8.28$\times$ 10$^{-5}$ cm$^{-3}$ and path length is L $\geqslant \frac{min(N_{H,2})^2}{max(EM_2)}$ = 352 kpc. Thus the hot phase is a low density medium, likely extended beyond the virial radius of the Galaxy.  

We should note that so far we have considered 3$\sigma$ confidence intervals of the temperatures. Considering a smaller (1$\sigma$) confidence interval, which has been used in earlier studies, we find that the temperature of the hot component in emission (T$^{em}_2$ = 10$^{6.68-6.92}$ K, including systematic uncertainty) and absorption (T$^{ab}_2$ = 10$^{6.96-7.15}$ K) do not overlap, although the difference is marginal (figure \ref{fig:T}). If the systems observed in emission and absorption are actually different, the calculation in the previous paragraph would not be applicable. Using the similar logic as discussed in \S\ref{sec:warmhot}, we can say that there may be multiple temperature components around 10$^7$K, which can be probed through a combined study of emission and absorption, using multiple elements heavier than oxygen as tracers. 

\subsection{Physical considerations}\label{sec:physics}
\noindent We summarize the discussion in \S\ref{compare:em} and \S\ref{compare:ab} in figure \ref{fig:T}. We confirm that there are two distinct phases in absorption \citep{Das2019a}, and also in emission (this work). The 10$^{6.1}$ K phase detected in absorption \citep{Das2019a} and in emission \citep{Kuntz2000} is clearly different from the 10$^{6.3-6.4}$ K phase, which is obtained by most of the emission and absorption studies \citep{Kuntz2000,Gupta2012,Henley2013,Gatuzz2018,Nakashima2018}. The hotter components have not been observed earlier in general, likely due to shallower data and/or poor spectral resolution. As the temperature of the hot absorbing phase is largely  uncertain, we are not sure whether the hot components observed in emission and absorption are the same. Therefore, the X-ray traced halo gas has at least 3 (or, 4) components spanning $\sim$2 orders of magnitude of temperature. While emission provides a strong constraint on the temperature, it is limited by the foreground uncertainty and the emissivity bias. Absorption, on the other hand, can probe multiple temperature components more efficiently, albeit with a weaker constraint. With emission-only or absorption-only measurements, we cannot determine whether the observed systems are co-spatial, and whether they reside in the Galactic halo, or beyond. 

The warm-hot component in emission (T$^{em}_1$) is the known and well-studied ambient halo gas at the virial temperature of the Galaxy, in hydrostatic equilibrium in the Galaxy's potential well. The difference between T$^{em}_1$ and T$^{abs}_1$ is small, so we cannot tell if they are genuinely different phases, or mere fluctuations, or different radial portions of a temperature gradient.  \citeauthor{Schmutzler1993},\citet{Gehrels1993} showed that the cooling curves of hot ($\geqslant 10^6$ K) optically thin plasma have local minima and maxima, whose positions are functions of metallicity, cooling process (isobaric or isochoric), ionization mechanism (collisional and/or photo-) and thermal history of the plasma. The gas is in a quasi-steady state at the maxima and thermally stable at the minima; so the gas accumulates at the minima while cooling. Thus, T$^{em}_1$ may correspond to the local maximum of cooling with higher emissivity, and T$^{abs}_1$ may correspond to the local minimum of cooling with accumulated gas. In that case, T$^{abs}_1$ would be a transition phase from the warm-hot (T$^{em}_1$) to the warm (T $< 10^6$ K) phase. 

The nature, position and hence the origin of the hot (T$^{em}_2$) component is not clear. Based on the temperature and emission measure arguments, we discuss the following possibilities: \\
I) If the hot system observed in emission and absorption are the same, it can reside within the halo and/or Local Group, or the hot intergalactic medium (IGM). Hydrodynamic simulations that take into account multistage stellar feedback from the bulges of MW-type galaxies predict that the temperature in the CGM can be as hot as T$^{em}_2$ \citep{Tang2009}. Analytic models predict that the CGM can significantly deviate from thermal equilibrium due to mechanical feedback (ejection of low angular momentum material) or thermal feedback (heating of the central regions), resulting in super-virial temperature as high as T$^{em}_2$ in the inner 50 kpc of the halo \citep{Pezzulli2017}. As the lower limit of the spatial extent of the hot gas (352 kpc) is larger than the virial radius R$_{200}$ of the Milky Way ($\sim$250 kpc), the hot gas may be present in the Local Group. T$^{em}_2$ is consistent with the predicted temperature of Local Group : 10$^{6.69-6.91}$ K \citep{Peebles1990}, allowing it to be in equilibrium. However, the sightline toward 1ES\,1553+113 ($l=21.91^\circ$, $b=43.96^\circ$) is away from M31 and the gravitational center of the Local Group; therefore the Local Group medium contribution to the T$^{em}_2$ component, if any, might not be significant. \cite{Oppenheimer2018} found that the thermal feedback can buoyantly rise to the outer CGM of MW-like halos of M$_{200} \leqslant 10^{12}$ M$_\odot$, moving baryons beyond R$_{200}$ and extending the CGM out to at least 2$\times$R$_{200}$. Whether to call this region the extended CGM or the Local Group medium, is just a matter of nomenclature. The observed hot gas can also come from the outskirt of the Local Group or IGM. This is substantiated by the fact that T$^{em}_2$ is consistent with the mass-weighted temperature of the baryons at z$\sim$0, and hot IGM has similar density as we obtain in \S\ref{sec:hot} \citep[and references therein]{Cen1999}. Due to the poor spectral resolution, we do not have any kinematic and redshift information of the hot gas - leaving the inference inconclusive. The mass of the hot ``shell" of gas will strongly depend on its distance from the Galactic center, as well as the covering factor and the volume-filling factor. In the most optimistic case of this component being ubiquitous, it would trace  a large amount of ``missing baryons", in the Galactic or cosmological scale. \\
II) As the sightline toward 1ES\,1553+113 passes close to the Fermi Bubble (FB): $|l|\leqslant 20^\circ$, $|b|\leqslant 50^\circ$ \citep{Su2010,Kataoka2018}, we investigate if the hot gas is from the structures around the FB. The hot absorbing gas is not related to the X-ray structures around FB \citep{Das2019a}, but the emitting gas might be, if the emitting and absorbing systems are not the same. A combined model of the halo gas, the FB and the X-ray shell around the FB based on \textit{Suzaku} and \xmm observations predicted a temperature of log($T_{shell}$/K) = 6.60--6.95, and density of n$_{shell}$ = 10$^{-3}$ cm$^{-3}$ \citep{Miller2016}. This temperature is similar with that of our hot emitting gas. The fiducial density provides a line-of-sight path length of EM$_2$/n$^2_{shell}$ = 1.8$^{+0.6}_{-0.9}$ kpc, consistent with the model of \cite{Miller2016}.  \cite{Kataoka2013, Kataoka2015} found a similar path length for the X-ray shell, but their inferred  temperature was $\approx$10$^{6.54}$ K. This is lower than the model temperature of \cite{Miller2016} and the temperature of our hot emission component. This might be an artifact of using a single temperature component to fit the soft X-ray emission, which might have yielded an average of the warm-hot CGM and the shell temperature. The hot gas can also be from the North Polar Spur (NPS). The temperature of NPS,  10$^{6.46-6.53}$ K \citep{Kataoka2018}, is in the valley between our warm-hot and hot component, and the emission measure of 0.02--0.07 cm$^{-6}$pc \citep{Kataoka2018} is comparable with the summation of our warm-hot and hot component (see \S\ref{sec:results}). This can, again, be an effect of 1-T modelling of the shallow observations. \\  

\section{Conclusion and future directions}\label{sec:conclusion}
\noindent We have studied the diffuse X-ray emission from the Milky Way halo toward 1ES\,1553+113, using deep \xmm observations. We have done a very detailed foreground analysis by using simple models usually used in literature, as well as more advanced models which have fewer assumptions and approximations. This has allowed us to determine the systematic uncertainties in the emission characteristics of the Galactic halo, dominating over the statistical uncertainties. Below, we summarize our science results: \\
1. We have found that a 2-temperature halo model better represents the data compared to a single temperature halo model. We refer to these as warm-hot (T$^{em}_1$ = 10$^{6.25-6.42}$ K) and hot (T$^{em}_2$ = 10$^{6.68-6.92}$ K) components. \\
2. The temperature of the warm-hot component is similar to most of the earlier measurements. However, it is not same as the temperature of the warm-hot component observed in absorption (T$^{ab}_1$) along the same sightline. This indicates the existence of multiple temperature components at and around the virial temperature.  \\ 
3. The hot component has never been detected as a prevalent component in emission, neither has it been associated with the Galactic halo before. Due to the large uncertainties in the temperature of the hot component in absorption (T$^{ab}_2$), we cannot determine if the emitting and the absorbing components are truly different. In the possibility of them being the same, we find that the hot component is a very low density (n$_{avg} \leqslant$ 8.28$\times$ 10$^{-5}$ cm$^{-3}$) gas extended beyond the virial radius (L $\geqslant$ 352 kpc) of the Galaxy. If they are different, the hot emitting component may be associated with the X-ray shell around the Fermi Bubble or the North Polar Spur.

The multi-component hot CGM, which we find in both absorption and emission along one sightline, may not be ubiquitous. To  characterize the multi-component hot halo gas between T$\sim$10$^{5.5-7.5}$ K, and to determine whether it is isotropic and homogeneous, it is essential to extend this study along as many sightlines as possible, with deep emission and absorption observations, and using multiple tracer elements like carbon, nitrogen, neon, magnesium and silicon in addition to oxygen. At present, the archival data of \chandra~ and \xmm can be very useful in this regard. On a longer timescale, planned missions like \textit{XRISM, Athena, Lynx} in the next decade and beyond will offer an outstanding opportunity to observe the highly ionized diffuse medium in unprecedented detail. This will bring us closer to understanding the co-evolution of the galaxy and its CGM. 
\section*{acknowledgement}
\noindent This work is based on observations obtained with \textit{XMM-Newton}, an ESA science mission with instruments and contributions directly funded by ESA Member States and NASA. S.M. acknowledges NASA grant NNX16AF49G. A.G. acknowledges support from the NASA ADAP grant 80NSSC18K0419. Y.K. acknowledges support from grant DGAPA-PAPIIT 106518, and from program DGAPA-PASPA.
\facilities{\xmm} 
\software{SAS v17.0.0 \citep{Snowden2004}, HeaSoft v6.17 \citep{Drake2005}, NumPy v1.11.0 \citep{Dubois1996}, Matplotlib v1.5.3 \citep{Hunter2007}} 
\appendix 
\section{Spectral Model}\label{sec:model}
\noindent As the instrumental components and the foreground of individual observations are not necessarily the same, we do not co-add the spectra. Instead, we simultaneously fit the 10 pn spectra within 0.33-7.0 keV in \texttt{XSPEC}\footnote{\url{https://heasarc.gsfc.nasa.gov/xanadu/xspec/manual/XspecManual.html}} using the following model components: \\
1. The \textbf{instrumental line} of Al K$\alpha$ at $\approx$1.48 keV is modeled as an unabsorbed zero-width Gaussian. The first 5 observations have similar line amplitudes, therefore they are tied to be the same. The last 5 observations also have similar line amplitudes, so they are also tied to each other. These two sets of observations have visibly different line amplitudes (figure \ref{fig:pn}), so they are allowed to be different. The \textbf{soft proton contamination} is modeled as an unabsorbed broken power-law. As suggested by ESAS manual, we keep the break-energy fixed at 3 keV, vary the power law indices between 0.1 to 10, with the response matrix not folded with the instrument. The power law indices and the normalization are kept free and are allowed to be different in all observations. \\
2. The most challenging part of the analysis is to model the \textbf{foreground}, whose uncertainties can strongly affect the derived properties of the Galactic halo. It is a combination of the Local hot bubble (LHB) and the Solar Wind Charge eXchange (SWCX), with their relative contribution varying with direction (for LHB and geocoronal SWCX) and time (for heliospheric SWCX). In extreme cases, the foreground is dominated by LHB or SWCX, where it is a reasonable approximation to model the foreground using just one component \citep{Henley2015b,Gupta2017}. Therefore, we have considered the following foreground models:
\\ \\
    a) \textbf{Model A} (LHB-only): We model LHB as an unabsorbed collisionally ionized plasma in thermal equilibrium (\texttt{apec}, see \cite{Das2019b} for details). We allow the temperature to vary in the range of k$_B$T = $0.097\pm0.013$ keV \citep{Liu2017}, freeze the metallicity at solar and keep the  normalization free. \\ \\
    b) SWCX-only models: We have used 2 SWCX models: \\
        I) \textbf{Model B} (AtomDB Charge eXchange code (\texttt{ACX}) based model): This model assumes that the charge exchange (CX) can be described as a thermal  emission, and it uses analytic expressions to calculate the distributions of the principal quantum number $n$ and the orbital angular momentum $l$ for the electron that transfers from the donor neutral atom to the receiving ion \citep{Smith2014}\footnote{the details with instructions are provided in \url{http://www.atomdb.org/CX/acx_manual.pdf}}. The relevant model parameters are temperature, normalization, He abundance, metallicity, and the distribution of $n$ and $l$. We freeze the He abundance and metallicity at cosmic and solar value, respectively. The parameter \texttt{model} has 16 options- the combinations of 2 $n$ distributions: a fixed $n$ or its weighted distribution, individual or total orbital angular momentum ($l$ or $L$), and 4 distributions of orbital angular momentum. \texttt{model}=7 and =8 use a weighted distribution of final $n$, and $l$ follows a distribution suitable for slow ($v < 1000$ km/s) solar winds. Therefore, we use \texttt{model}=8 in our analysis (\texttt{model}=7 also gave consistent results). The temperature and normalization are kept free. \\
        II) \textbf{Model C} (Line-ratio constrained model): In this model we do not assume CX as a thermal emission, but the level populations are assumed to follow the distribution in thermal equilibrium. This is similar to the SWCX model of \cite{Henley2015a} for \cvi Ly$\alpha-\delta$, \ovii K$\alpha (f+i+r)-\epsilon$ and \oviii Ly$\alpha-\epsilon$ line emissions. We add \neix K$\alpha-\gamma$, \nex Ly$\alpha-\beta$ and \mgxi K$\alpha$ lines and freeze the line ratios to  \cite{Cumbee2014}.\\ \\
    c) LHB+SWCX: The first three models (A, B, C) assume that the foreground is dominated by either SWCX or LHB. However, their contributions may in fact be comparable \citep{Henley2015b}. Therefore, we consider 3 more foreground models: \\
        I) \textbf{Model D} (LHB+\texttt{ACX}) and II) \textbf{model E} (LHB+Line-ratio constrained SWCX): The emission measure of SWCX-subtracted LHB is allowed to vary within 0.8-6.5 $\times$10$^{-3}$ cm$^{-6}$ pc \citep{Liu2017} in both models. The temperature and metallicity of LHB are varied/frozen the same as in model A. \\
        III) \textbf{Model F} (LHB+unconstrained CX): As the CX is a non-thermal and not necessarily an equilibrium process, the ratio between different line transitions of the same ion do not need to have fixed values. We added eight lines at the positions of \cvi Ly$\alpha$, \cvi Ly$\beta$, \ovii K$\alpha (f+i+r)$, \oviii Ly$\alpha$+\ovii K$\beta-\epsilon$, \oviii Ly$\beta-\epsilon$, \neix K$\alpha$, \nex Ly$\alpha$ and \mgxi K$\alpha$. The spectral resolution of pn CCDs cannot resolve most of these lines, so we model them as zero-width Gaussians. The energy of single lines are kept fixed and the energies of composite lines are allowed to vary within the range of central energies of overlapping lines. This is the most general model where we do not make any assumption about the foreground. This is the first time soft X-ray foreground is modeled as a combination of LHB and SWCX, without assuming either of these to have a negligible contribution.\\
3. The \textbf{cosmic X-ray background} (CXB) due to unresolved point sources is modeled as a power law, absorbed by the Galactic interstellar medium [\texttt{phabs*powerlaw}]. We keep the normalization and the power law index as  free parameters. We keep the absorbing column density N(\hi) along this sightline fixed at 3.92$\times$10$^{20}$ cm$^{-2}$ \citep{Gatuzz2018}. Using Tuebingen-Boulder ISM absorption model (\texttt{tbabs}) instead of \texttt{phabs} made no difference. \\
 4. With our prior knowledge of two-temperature CGM discovered in absorption studies along this sightline \citep{Das2019a}, we model the \textbf{Galactic halo} as an absorbed two-temperature collisionally ionized plasma in thermal equilibrium [\texttt{phabs*(apec+apec)}]. We freeze the metallicity at solar. The normalization factor of the thermal plasma model is metallicity-weighted, so the exact value of the input metallicity does not matter. 

The foreground, background and the Galactic halo of all observations are tied to be the same. Although SWCX can vary with time, allowing it to be different in all observations did not produce any appreciable difference.
\\ 
We take into account thermal broadening of the lines and the pseudo-continuum (low-flux lines which are not individually stored in the AtomDB output files) by switching \texttt{apecthermal} and \texttt{apecbroadpseudo} on, respectively. The chemical composition of the \texttt{ACX} and \texttt{apec} models have been set to solar according to the prescription of \cite{Asplund2009}. Using other prescriptions \citep{Wilms2000,Lodders2003} did not result in noticeable differences. This is expected, because the emission spectrum is dominated by oxygen lines, and the prescriptions we have used have similar oxygen abundances. The difference in C/O, N/O, Ne/O and O/Fe ratios do not affect the spectrum appreciably. 
\begin{table}[]
\centering
\caption{\textcolor{black}{Best fit parameters of the models (foreground+background)}}
\label{tab:CEF}
\textcolor{black}{
\begin{tabular}{@{}c|cc|cccccc@{}}
\toprule
Model &  &  & A & B & C & D & E & F \\\midrule
\multirow{2}{*}{LHB} & kT & (keV) & 0.11 & -- & -- & 0.11 & 0.11 & 0.11 \\
 & norm & (cm$^{-5}$) & 1.2$\times$10$^{-3}$ & -- & -- & 6.5$\times$10$^{-4}$ & 6.5$\times$10$^{-4}$ & 8.2$\times$10$^{-4}$ \\\hline
\multirow{11}{*}{SWCX} & {ACX}(kT) & (keV) & \multirow{11}{*}{--} & 0.15 & \multirow{2}{*}{--} & 0.18 & \multirow{2}{*}{--} & \multirow{2}{*}{--} \\
 & {ACX}(norm) & (cm$^{-5}$) &  & 1.22$\times$10$^{-5}$ &  & 1.30$\times$10$^{-5}$ &  &  \\
 & \cvi Ly$\alpha$ & \multirow{9}{*}{(cnt s$^{-1}$cm$^{-2}$)} &  & \multirow{9}{*}{--} & 8.6$\times$10$^{-4}$ & \multirow{9}{*}{--} & 6.7$\times$10$^{-4}$ & 3.7$\times$10$^{-4}$ \\
 & \cvi Ly$\beta$ &  &  &  & -- &  & -- & 1.9$\times$10$^{-4}$ \\
 & \ovii K$\alpha(f+i+r)$ &  &  &  & 1.5$\times$10$^{-4}$ &  & 1.4$\times$10$^{-4}$ & $\approx$0 \\
 & \ovii K$\beta$+\oviii Ly$\alpha$ &  &  &  & -- &  & -- & 6.7$\times$10$^{-5}$ \\
 & \oviii Ly$\alpha$ &  &  &  & 4.1$\times$10$^{-5}$ &  & 1.2$\times$10$^{-4}$ & -- \\
 & \oviii Ly$\beta-\epsilon$ &  &  &  & -- &  & -- & 2.7$\times$10$^{-5}$ \\
 & \neix  K$\alpha$ &  &  &  & 1.5$\times$10$^{-5}$ &  & 3.3$\times$10$^{-5}$ & 3.4$\times$10$^{-5}$ \\
 & \nex Ly$\alpha$ &  &  &  & 2.6$\times$10$^{-6}$ &  & 3.8$\times$10$^{-6}$ & 3.6$\times$10$^{-6}$ \\
 & \mgxi K$\alpha$ &  &  &  & 3.3$\times$10$^{-6}$ &  & 2.9$\times$10$^{-6}$ & 3.6$\times$10$^{-6}$\\\hline
 \multirow{2}{*}{CXB} & $\Gamma$ &  & 1.9 & 1.7 & 1.8 & 1.7 & 2.0 & 1.8 \\
 & norm & (cnt s$^{-1}$cm$^{-2}$keV$^{-1}$) & 3.4$\times$10$^{-4}$ & 3.8$\times$10$^{-4}$ & 4.4$\times$10$^{-4}$ & 4.3$\times$10$^{-4}$ & 4.7$\times$10$^{-4}$ & 4.4$\times$10$^{-4}$ \\\hline
\end{tabular}%
}
\end{table}
\section{Is the hot component real?}
\noindent Using absorption studies, \cite{Das2019a} found a two-temperature model for the MW CGM, with both warm-hot and hot components. This does not necessarily require that we find the hot component in the emission spectrum. Therefore, in the context of multi-temperature components of highly ionized CGM, it is necessary to test if the emission spectrum really needs two components for the Galactic halo. As the hot halo is an additive component, we can verify its presence using the F-test. Spectral models with foreground models A, B or C are not suitable for this purpose, because the foreground is approximated to one extreme (only LHB or only SWCX). In model E and F, the foreground has Gaussian lines at the energies where the hot halo is dominant; they are not suited for the F-test either. So, we perform the F-test on the model with foreground model D, which has both LHB and SWCX. The F statistic value, for \textcolor{black}{$\chi_{\nu,1T}^2$ = 1.238976 and $\chi_{\nu,2T}^2$ = 1.046320}, is 456.391 with null hypothesis probability $\sim$ 0. This shows that we do need the hot halo component with extremely high confidence. 
\section{Sanity check with MOS2}\label{sec:mos}
\begin{figure}
    \centering
    \includegraphics[scale=0.425]{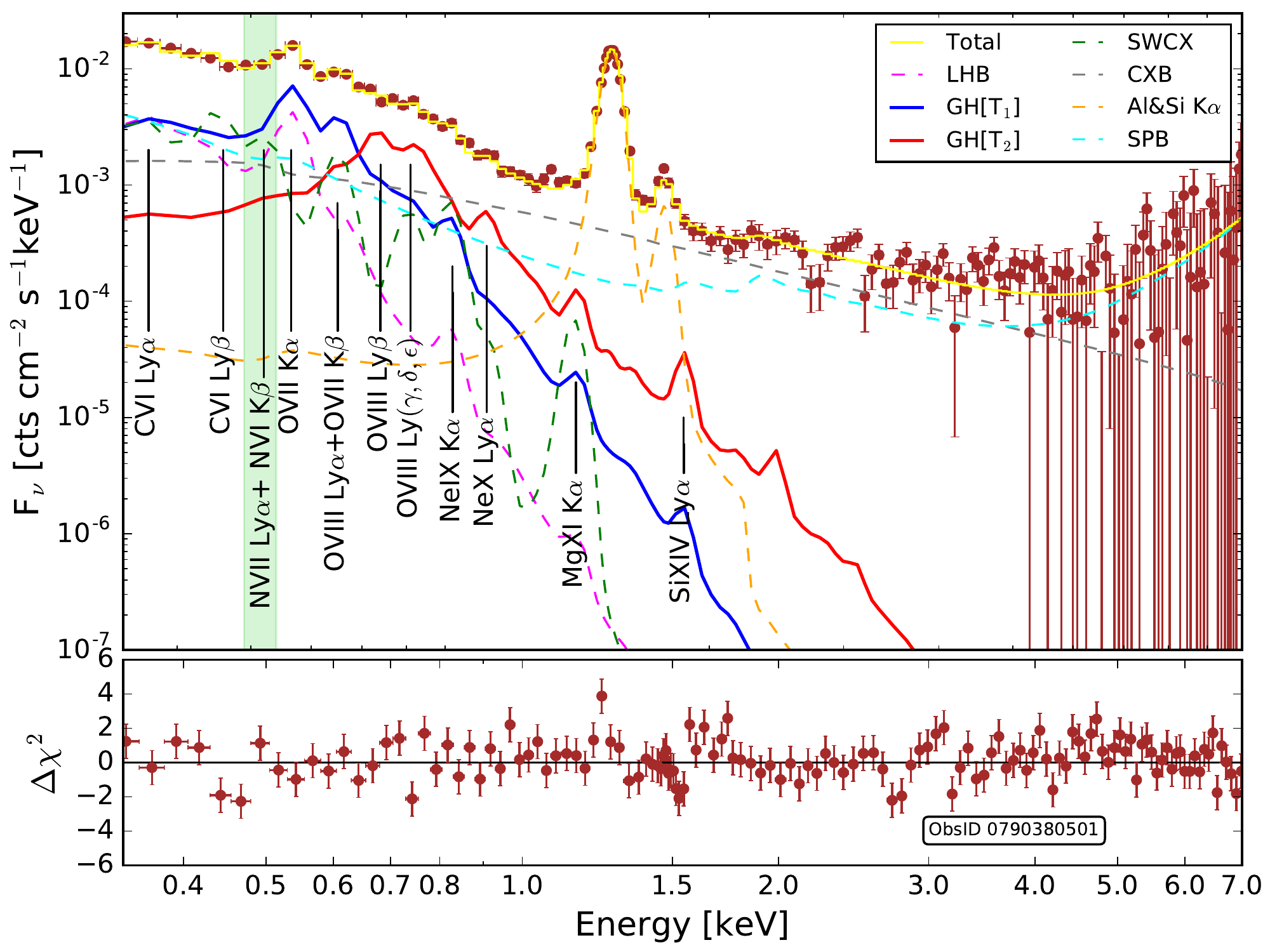}
    \caption{One MOS2 spectrum decomposed into the instrumental lines, foreground (model F), background and the halo (see appendix \ref{sec:model} for the details of the legend). The characteristic emission lines of carbon, oxygen, neon, magnesium and silicon have been labeled, so that the relative contribution of the spectral components at each line can be studied independently. As MOS2 has better spectral resolution compared to pn, the characteristic emission lines of the two halo components are more evident in the MOS spectrum compared to those in pn (figure \ref{fig:pn}). For example, the warm-hot (shown in blue) system emits most of the \ovii K$\alpha-\beta$ and \oviii Ly$\alpha$ around 0.6 keV, while most of the \oviii Ly$\beta$ and Ly$\gamma-\epsilon$ emission around 0.8 keV come from the the hot (shown in red) phase.}
    \label{fig:mos}
\end{figure}

The effective area of MOS is much smaller than that of pn, and the instrumental background is higher than pn, therefore we performed our data analysis on the pn data only. However, we use the MOS data to check for consistency with the EPIC-pn data. 
MOS has better spectral resolution, therefore we cannot fit the spectra of pn and MOS independently and  compare the two independent results, especially when the CX model includes Gaussian lines at fixed energies. Because of different spectral resolutions, the temperature estimates from pn and MOS2 spectra would not necessarily be the same either. Therefore, we fit the 10 MOS2 spectra in the range of 0.325-10 keV keeping the non-instrumental spectral components fixed at their best-fitted values of the 10 pn spectra. In addition to the spectral model described in \S\ref{sec:model}, we add a Gaussian for the instrumental Si K$\alpha$ line at 1.74 keV. The soft proton background is independently fitted. To account for the uncertainty due to inter-instrumental calibration, we scale the non-instrumental spectral components with a constant factor that is allowed to vary between 0.9 and 1.1. We find that the best-fitted model for the pn spectra does not fit the MOS2 spectra well ($\chi^2/dof$ = 13281.95/11624; foreground model D), due to excess emission around 0.515 keV. Changing the default abundance ratio \citep{Asplund2009} to different prescriptions \citep{Wilms2000,Lodders2003} could not account for this excess. This excess was not visible in the pn spectra, because of the lower spectral resolution. \nvii Ly-$\alpha$ and/or \nvi K$\beta$ lines are the only feasible sources of the excess emission. This excess emission from nitrogen can arise due to two different reasons: 1) the N/O ratio of the foreground (LHB) and/or the Galactic halo components are super-solar; or 2) this is a SWCX emission. The good spectral resolution of MOS2 and the better S/N of the spectra compared to earlier studies have allowed us to distinguish the nitrogen line from its surrounding emission. After including the \nvii/\nvi line in the model, the fit improves ($\chi^2/dof$ = 12856.35/11621; foreground model D), with the inter-instrumental  calibration factor of $\approx$1.05 (figure \ref{fig:mos}). This shows that similar models with the same halo components fit both pn and MOS spectra.


\begin{thebibliography}{}
\expandafter\ifx\csname natexlab\endcsname\relax\def\natexlab#1{#1}\fi
\providecommand{\url}[1]{\href{#1}{#1}}

\bibitem[{{Asplund} {et~al.}(2009){Asplund}, {Grevesse}, {Sauval}, \&
  {Scott}}]{Asplund2009}
{Asplund}, M., {Grevesse}, N., {Sauval}, A.~J., \& {Scott}, P. 2009, \araa, 47,
  481

\bibitem[{{Cen} \& {Ostriker}(1999)}]{Cen1999}
{Cen}, R., \& {Ostriker}, J.~P. 1999, \apj, 514, 1

\bibitem[{{Cumbee} {et~al.}(2014){Cumbee}, {Henley}, {Stancil}, {Shelton},
  {Nolte}, {Wu}, \& {Schultz}}]{Cumbee2014}
{Cumbee}, R.~S., {Henley}, D.~B., {Stancil}, P.~C., {et~al.} 2014, \apjl, 787,
  L31

\bibitem[{{Das} {et~al.}(2019{\natexlab{a}}){Das}, {Mathur}, {Gupta},
  {Nicastro}, {Krongold}, \& {Null}}]{Das2019b}
{Das}, S., {Mathur}, S., {Gupta}, A., {et~al.} 2019{\natexlab{a}}, ApJ, in
  press, arXiv:1810.12454

\bibitem[{{Das} {et~al.}(2019{\natexlab{b}}){Das}, {Mathur}, {Nicastro}, \&
  {Krongold}}]{Das2019a}
{Das}, S., {Mathur}, S., {Nicastro}, F., \& {Krongold}, Y. 2019{\natexlab{b}},
  \apjl, 882, L23

\bibitem[{{Dav{\'e}} {et~al.}(2010){Dav{\'e}}, {Oppenheimer}, {Katz},
  {Kollmeier}, \& {Weinberg}}]{Dave2010}
{Dav{\'e}}, R., {Oppenheimer}, B.~D., {Katz}, N., {Kollmeier}, J.~A., \&
  {Weinberg}, D.~H. 2010, \mnras, 408, 2051

\bibitem[{{Drake}(2005)}]{Drake2005}
{Drake}, S.~A. 2005, in X-Ray and Radio Connections, ed. L.~O. {Sjouwerman} \&
  K.~K. {Dyer} (Santa Fe, New Mexico: NRAO), 6.01

\bibitem[{{Dubois} {et~al.}(1996){Dubois}, {Hinsen}, \& {Hugunin}}]{Dubois1996}
{Dubois}, P.~F., {Hinsen}, K., \& {Hugunin}, J. 1996, Computers in Physics, 10,
  262

\bibitem[{{Ford} {et~al.}(2014){Ford}, {Dav{\'e}}, {Oppenheimer}, {Katz},
  {Kollmeier}, {Thompson}, \& {Weinberg}}]{Ford2014}
{Ford}, A.~B., {Dav{\'e}}, R., {Oppenheimer}, B.~D., {et~al.} 2014, \mnras,
  444, 1260

\bibitem[{{Gatuzz} \& {Churazov}(2018)}]{Gatuzz2018}
{Gatuzz}, E., \& {Churazov}, E. 2018, \mnras, 474, 696

\bibitem[{{Gehrels} \& {Williams}(1993)}]{Gehrels1993}
{Gehrels}, N., \& {Williams}, E.~D. 1993, \apjl, 418, L25

\bibitem[{{Gupta} {et~al.}(2014){Gupta}, {Mathur}, {Galeazzi}, \&
  {Krongold}}]{Gupta2014}
{Gupta}, A., {Mathur}, S., {Galeazzi}, M., \& {Krongold}, Y. 2014, \apss, 352,
  775

\bibitem[{{Gupta} {et~al.}(2017){Gupta}, {Mathur}, \& {Krongold}}]{Gupta2017}
{Gupta}, A., {Mathur}, S., \& {Krongold}, Y. 2017, \apj, 836, 243

\bibitem[{{Gupta} {et~al.}(2012){Gupta}, {Mathur}, {Krongold}, {Nicastro}, \&
  {Galeazzi}}]{Gupta2012}
{Gupta}, A., {Mathur}, S., {Krongold}, Y., {Nicastro}, F., \& {Galeazzi}, M.
  2012, \apjl, 756, L8

\bibitem[{{Henley} \& {Shelton}(2013)}]{Henley2013}
{Henley}, D.~B., \& {Shelton}, R.~L. 2013, \apj, 773, 92

\bibitem[{{Henley} \& {Shelton}(2015)}]{Henley2015b}
---. 2015, \apj, 808, 22

\bibitem[{{Henley} {et~al.}(2015){Henley}, {Shelton}, {Cumbee}, \&
  {Stancil}}]{Henley2015a}
{Henley}, D.~B., {Shelton}, R.~L., {Cumbee}, R.~S., \& {Stancil}, P.~C. 2015,
  \apj, 799, 117

\bibitem[{{Henley} {et~al.}(2010){Henley}, {Shelton}, {Kwak}, {Joung}, \& {Mac
  Low}}]{Henley2010}
{Henley}, D.~B., {Shelton}, R.~L., {Kwak}, K., {Joung}, M.~R., \& {Mac Low},
  M.-M. 2010, \apj, 723, 935

\bibitem[{{Hunter}(2007)}]{Hunter2007}
{Hunter}, J.~D. 2007, Computing in Science and Engineering, 9, 90

\bibitem[{{Kataoka} {et~al.}(2018){Kataoka}, {Sofue}, {Inoue}, {Akita},
  {Nakashima}, \& {Totani}}]{Kataoka2018}
{Kataoka}, J., {Sofue}, Y., {Inoue}, Y., {et~al.} 2018, Galaxies, 6, 27

\bibitem[{{Kataoka} {et~al.}(2015){Kataoka}, {Tahara}, {Totani}, {Sofue},
  {Inoue}, {Nakashima}, \& {Cheung}}]{Kataoka2015}
{Kataoka}, J., {Tahara}, M., {Totani}, T., {et~al.} 2015, \apj, 807, 77

\bibitem[{{Kataoka} {et~al.}(2013){Kataoka}, {Tahara}, {Totani}, {Sofue},
  {Stawarz}, {Takahashi}, {Takeuchi}, {Tsunemi}, {Kimura}, {Takei}, {Cheung},
  {Inoue}, \& {Nakamori}}]{Kataoka2013}
---. 2013, \apj, 779, 57

\bibitem[{{Kuntz} \& {Snowden}(2000)}]{Kuntz2000}
{Kuntz}, K.~D., \& {Snowden}, S.~L. 2000, \apj, 543, 195

\bibitem[{{Liu} {et~al.}(2017){Liu}, {Chiao}, {Collier}, {Cravens}, {Galeazzi},
  {Koutroumpa}, {Kuntz}, {Lallement}, {Lepri}, {McCammon}, {Morgan}, {Porter},
  {Snowden}, {Thomas}, {Uprety}, {Ursino}, \& {Walsh}}]{Liu2017}
{Liu}, W., {Chiao}, M., {Collier}, M.~R., {et~al.} 2017, \apj, 834, 33

\bibitem[{{Lodders}(2003)}]{Lodders2003}
{Lodders}, K. 2003, \apj, 591, 1220

\bibitem[{{Miller} \& {Bregman}(2016)}]{Miller2016}
{Miller}, M.~J., \& {Bregman}, J.~N. 2016, \apj, 829, 9

\bibitem[{{Nakashima} {et~al.}(2018){Nakashima}, {Inoue}, {Yamasaki}, {Sofue},
  {Kataoka}, \& {Sakai}}]{Nakashima2018}
{Nakashima}, S., {Inoue}, Y., {Yamasaki}, N., {et~al.} 2018, \apj, 862, 34

\bibitem[{{Nevalainen} {et~al.}(2017){Nevalainen}, {Wakker}, {Kaastra},
  {Bonamente}, {Snowden}, {Paerels}, \& {de Vries}}]{Nevalainen2017}
{Nevalainen}, J., {Wakker}, B., {Kaastra}, J., {et~al.} 2017, \aap, 605, A47

\bibitem[{{Nicastro} {et~al.}(2016a){Nicastro}, {Senatore}, {Gupta},
  {Guainazzi}, {Mathur}, {Krongold}, {Elvis}, \& {Piro}}]{Nicastro2016a}
{Nicastro}, F., {Senatore}, F., {Gupta}, A., {et~al.} 2016a, \mnras, 457, 676

\bibitem[{{Nicastro} {et~al.}(2016b){Nicastro}, {Senatore}, {Krongold},
  {Mathur}, \& {Elvis}}]{Nicastro2016b}
{Nicastro}, F., {Senatore}, F., {Krongold}, Y., {Mathur}, S., \& {Elvis}, M.
  2016b, \apj, 828, L12

\bibitem[{{Nicastro} {et~al.}(2018){Nicastro}, {Kaastra}, {Krongold},
  {Borgani}, {Branchini}, {Cen}, {Dadina}, {Danforth}, {Elvis}, \&
  {Fiore}}]{Nicastro2018}
{Nicastro}, F., {Kaastra}, J., {Krongold}, Y., {et~al.} 2018, \nat, 558, 406

\bibitem[{{Oppenheimer}(2018)}]{Oppenheimer2018}
{Oppenheimer}, B.~D. 2018, \mnras, 480, 2963

\bibitem[{{Oppenheimer} {et~al.}(2016){Oppenheimer}, {Crain}, {Schaye},
  {Rahmati}, {Richings}, {Trayford}, {Tumlinson}, {Bower}, {Schaller}, \&
  {Theuns}}]{Oppenheimer2016}
{Oppenheimer}, B.~D., {Crain}, R.~A., {Schaye}, J., {et~al.} 2016, \mnras, 460,
  2157

\bibitem[{{Peebles}(1990)}]{Peebles1990}
{Peebles}, P.~J.~E. 1990, \apj, 362, 1

\bibitem[{{Peeples} {et~al.}(2014){Peeples}, {Werk}, {Tumlinson},
  {Oppenheimer}, {Prochaska}, {Katz}, \& {Weinberg}}]{Peeples2014}
{Peeples}, M.~S., {Werk}, J.~K., {Tumlinson}, J., {et~al.} 2014, \apj, 786, 54

\bibitem[{{Pezzulli} {et~al.}(2017){Pezzulli}, {Fraternali}, \&
  {Binney}}]{Pezzulli2017}
{Pezzulli}, G., {Fraternali}, F., \& {Binney}, J. 2017, \mnras, 467, 311

\bibitem[{{Putman} {et~al.}(2012){Putman}, {Peek}, \& {Joung}}]{Putman2012}
{Putman}, M.~E., {Peek}, J.~E.~G., \& {Joung}, M.~R. 2012, \araa, 50, 491

\bibitem[{{Schmutzler} \& {Tscharnuter}(1993)}]{Schmutzler1993}
{Schmutzler}, T., \& {Tscharnuter}, W.~M. 1993, \aap, 273, 318

\bibitem[{Smith {et~al.}(2014)Smith, Foster, Edgar, \& Brickhouse}]{Smith2014}
Smith, R.~K., Foster, A.~R., Edgar, R.~J., \& Brickhouse, N.~S. 2014, \apj,
  787, 77

\bibitem[{{Snowden} {et~al.}(2004){Snowden}, {Valencic}, {Perry}, {Arida}, \&
  {Kuntz}}]{Snowden2004}
{Snowden}, S., {Valencic}, L., {Perry}, B., {Arida}, M., \& {Kuntz}, K.~D.
  2004, {The XMM-Newton ABC Guide: An Introduction to XMM-Newton Data
  Analysis}, Tech. rep.,
  doi:https://heasarc.gsfc.nasa.gov/docs/xmm/abc/abc.html

\bibitem[{{Snowden} {et~al.}(2000){Snowden}, {Freyberg}, {Kuntz}, \&
  {Sanders}}]{Snowden2000}
{Snowden}, S.~L., {Freyberg}, M.~J., {Kuntz}, K.~D., \& {Sanders}, W.~T. 2000,
  \apjs, 128, 171

\bibitem[{{Spitzer}(1956)}]{Spitzer1956}
{Spitzer}, Lyman, J. 1956, \apj, 124, 20

\bibitem[{{Stinson} {et~al.}(2012){Stinson}, {Brook}, {Prochaska}, {Hennawi},
  {Shen}, {Wadsley}, {Pontzen}, {Couchman}, {Quinn}, \&
  {Macci{\`o}}}]{Stinson2012}
{Stinson}, G.~S., {Brook}, C., {Prochaska}, J.~X., {et~al.} 2012, \mnras, 425,
  1270

\bibitem[{{Su} {et~al.}(2010){Su}, {Slatyer}, \& {Finkbeiner}}]{Su2010}
{Su}, M., {Slatyer}, T.~R., \& {Finkbeiner}, D.~P. 2010, \apj, 724, 1044

\bibitem[{{Suresh} {et~al.}(2017){Suresh}, {Rubin}, {Kannan}, {Werk},
  {Hernquist}, \& {Vogelsberger}}]{Suresh2017}
{Suresh}, J., {Rubin}, K. H.~R., {Kannan}, R., {et~al.} 2017, \mnras, 465, 2966

\bibitem[{{Tang} {et~al.}(2009){Tang}, {Wang}, {Lu}, \& {Mo}}]{Tang2009}
{Tang}, S., {Wang}, Q.~D., {Lu}, Y., \& {Mo}, H.~J. 2009, \mnras, 392, 77

\bibitem[{{Tumlinson} {et~al.}(2017){Tumlinson}, {Peeples}, \&
  {Werk}}]{Tumlinson2017}
{Tumlinson}, J., {Peeples}, M.~S., \& {Werk}, J.~K. 2017, \araa, 55, 389

\bibitem[{{Voit} {et~al.}(2015){Voit}, {Donahue}, {Bryan}, \&
  {McDonald}}]{Voit2015}
{Voit}, G.~M., {Donahue}, M., {Bryan}, G.~L., \& {McDonald}, M. 2015, \nat,
  519, 203

\bibitem[{{Wilms} {et~al.}(2000){Wilms}, {Allen}, \& {McCray}}]{Wilms2000}
{Wilms}, J., {Allen}, A., \& {McCray}, R. 2000, \apj, 542, 914

\bibitem[{{Yao} {et~al.}(2009a){Yao}, {Wang}, {Hagihara}, {Mitsuda},
  {McCammon}, \& {Yamasaki}}]{Yao2009a}
{Yao}, Y., {Wang}, Q.~D., {Hagihara}, T., {et~al.} 2009a, \apj, 690, 143

\end{thebibliography}

\end{document}